\documentclass[fleqn,usenatbib]{mnras}

\usepackage{newtxtext,newtxmath}

\usepackage[T1]{fontenc}

\DeclareRobustCommand{\VAN}[3]{#2}
\let\VANthebibliography\thebibliography
\def\thebibliography{\DeclareRobustCommand{\VAN}[3]{##3}\VANthebibliography}

\usepackage{graphicx}	
\usepackage{amsmath}	
\usepackage{xcolor}

\title[The survivability of VLA1623]{Exploring the survivability of higher-order multiple protostellar systems -- The case of VLA1623}

\author[Murillo \& P\'erez-Villegas]{
	N. M. Murillo,$^{1}$\thanks{E-mail: nmurillo@astro.unam.mx}
	and A. P\'erez-Villegas,$^{1}$
	\\
	$^{1}$Universidad Nacional Autónoma de México, Instituto de Astronom\'ia, A.P. 106,  22800, Ensenada, B.C., México
}

\date{Accepted XXX. Received YYY; in original form ZZZ}

\pubyear{2025}

\begin{document}
	\label{firstpage}
	\pagerange{\pageref{firstpage}--\pageref{lastpage}}
	\maketitle
	
	\begin{abstract}

        Higher-order protostellar systems ($\geq$3 components) are commonly observed in the early stages of low-mass star formation. A persistent question in star formation and evolution is whether these higher-order protostellar systems survive as gravitationally bound systems or dissolve into binaries over time. We explore this question with a case study of the embedded quadruple protostellar system VLA1623 and its observational constraints, assuming that the components A1, A2, B, and W are gravitationally bound.
        Using $N$-body simulations, we run a grid of models considering gravity, mass accretion onto protostars, the presence of the protostellar cloud core, and its dispersal. The simulations are integrated over a period of 8 Myr to take into account the evolution from the protostellar phase (Class 0 and I, 1 Myr), through the pre-main sequence (Class II and III, 2 -- 3 Myr), and into the main sequence (4 Myr). Our results show that VLA1623 has a probability of 30\% to remain as a gravitationally-bound quadruple system up to 8 Myr from the current state. 
        There is also a 25--30\% probability of VLA1623 dissolving into a triple system. This suggests that the dissolution of quadruple protostellar systems contributes to the formation of stable triple (proto)stellar systems.
        Components A1 and A2 are the most likely to be ejected, while W has a lower probability of being ejected from the system. The stability of VLA1623 depends on a combination of component mass ratios wrt the primary, separations, and eccentricities.

	\end{abstract}
	
	\begin{keywords}
		stars: protostars -- stars: binaries -- stars: kinematics and dynamics -- methods: numerical
	\end{keywords}
	
	
	
	\section{Introduction}
	\label{sec:intro}

    During the star formation process, multiplicity is a common outcome and its occurrence increases with mass (\citealt{offner2023} and references therein). At the earliest stages of the star formation process, dust continuum surveys in the low-mass regime have identified significant multiplicity and companion fractions within low-mass protostellar cloud cores \citep{looney2000,chen2013,tobin2016,tobin2022VANDAM_Ori}. Stable higher-order protostellar systems ($\geq$3 gravitationally-bound stars) have been identified in the pre-main sequence stage (e.g., \citealt{reipurth2024}), and are ubiquitous and central to a wide range of phenomena in our universe (e.g.,\citealt{safarzadeh2020,toonen2020,hamers2020,lagos2024,shariat20253to2,dorozsmai2025,shariat2025CVs,kummer2025}). The survivability rate of higher-order systems as gravitationally-bound systems from the embedded phase into the main sequence is thus an open question.

The formation and evolution of a protostellar system is typically modeled using (non-ideal magneto-) hydrodynamical simulations \citep[e.g.,][]{lee2019,offner2023,kuruwita2023,mignonrisse2023,offner2025} due to the significant presence of gas, dust, and magnetic fields during the process. However, the computational cost of these simulations limits the exploration of a broad parameter space and the long integration times needed to study the survivability of higher-order systems from the protostellar phase to the main sequence stage. Therefore, an alternative approach is required to address the question of survivability of higher-order protostellar systems. The survival of protostellar and pre-main sequence triples has been studied using N-body simulations, occasionally incorporating smooth particle hydrodynamics \citep[SPH; e.g.,][]{sterzik1998,reipurth2000,reipurth2010,umbreit2011,reipurth2012,reipurth2015,rawiraswattana2012}. This method allows for the exploration of the dynamical evolution of higher-order protostellar systems, although it sacrifices a detailed treatment of gas, dust, and magnetic fields. The effects of the protostellar cloud core, mass accretion onto the protostars, and mass loss from the cloud core can be simulated by adjusting the gravitational potential, as demonstrated by \citet{reipurth2015}, who approximated the cloud core with a Plummer sphere and calculated accretion using a Bondi-Hoyle prescription.

Much of the previous observations and theory have focused on binary and triple systems at all stages of stellar evolution (e.g., \citealt{eggleton1995,reipurth2000,umbreit2005,reipurth2010,umbreit2011,rawiraswattana2012,korntreff2012,reipurth2012,reipurth2015,tokovininv2020formation,toonen2020,elbadry2021,hamers2022,toonen2022,lagos2024,shariat20253to2,dorozsmai2025,shariat2025CVs,shariat2025gaia,kummer2025}, see also the PPVII review \citealt{offner2023} and references therein). 
In parallel, most studies have focused on small separation binaries and triples, with inclusion of wider separations being a more recent development.
Much less attention has been paid to higher-order (proto)stellar systems (e.g., \citealt{sterzik1998,pineda2015,safarzadeh2020,hamers2020,han2022,kostov2022}), and quadruples in particular.
This makes sense considering the statistics of main sequence multiple stars which find the number of triples and quadruples to be two and four orders of magnitude lower, respectively, than binaries \citep{elbadry2021,shariat2025gaia}.
Despite their smaller numbers, higher-order stellar systems are still significant, both from a statistical and dynamical perspective.

Quadruples have been identified in stable systems with a range of mass ratios, orbital periods, eccentricities, and configurations (e.g., \citealt{tokovinin2006,tokovinin2020quadobs,volkov2021,han2022,kostov2022,reipurth2024,powell2025}). 
They can survive well into the late stages of stellar evolution to produce exotic phenomena (e.g., \citealt{hamers2020,safarzadeh2020}), and even survive stellar death (e.g., \citealt{demarco2022}).
In star forming regions, it is not uncommon to find four or more protostellar objects forming within a single cloud core (e.g., \citealt{tobin2016,tobin2022VANDAM_Ori}).
It is not well understood whether protostars that are observed to be in a common cloud core are and will remain gravitationally-bound. Sparse attempts have been made to address this question (e.g., L1448 IRS3 A and B: \citealt{kwon2006}), but it is still an open question.

VLA 1623-2417 (hereafter VLA1623), located in L1688 at a distance of 138 $\pm$ 2.6 pc \citep{ortiz2018}, is one of the most widely studied embedded quadruple protostellar systems (A1+A2\footnote{These components are also referred to in literature as Aa+Ab \citep{radley2025,sadavoy2018,sadavoy2024vla1623masses}. In the present work, we adopt A1+A2 as it is more common in embedded protostellar systems, and the lower case nomenclature is more commonly associated with exoplanets during the star formation process.}, B, and W) with three rotationally-supported gas disks \citep[e.g.,][]{andre1990,murillo2013,harris2018A,kawabe2018,ohashi2022,codella2022,mercimek2023,codella2024,sadavoy2024vla1623masses,radley2025,mercimek2025,maureira2025}.
Analysis of the rotationally-supported disks enabled protostellar masses to be derived for components A1+A2, B and W \citep{ohashi2022,sadavoy2024vla1623masses}, making VLA1623 an ideal case study.

Recently, \citet{mercimek2023} proposed that component W is gravitationally unbound based on the projected spatial separation and velocity offsets wrt A+B. The authors proposed two scenarios to explain this: (1) W and A+B formed in separate cores, or (2) W was ejected from the system during early formation. Similar scenarios were explored earlier in \citet{murillo2013}. In contrast, \citet{sadavoy2024vla1623masses} favor W as part of the VLA1623 system given their proper motion analysis and the presence of gas streamers connecting A+B and W (e.g., \citealt{mercimek2025}). However, the available data cannot unambiguously determine the membership of W. Consequently, this paper uses observational constraints of VLA1623 coupled with N-body simulations to explore its survivability as a gravitationally bound system.

The paper is structured as follows.
Sect.~\ref{sec:VLA1623data} describes the VLA1623 system in detail, including all adopted observational constraints used in the simulations.
The $N$-body simulations, set-up, parameter space, and diagnostics will be detailed in Sect.~\ref{sec:simulations}.
The outcome of the simulations will be discussed in Sect.~\ref{sec:results}. 
Discussion and conclusions will be left to Sect.~\ref{sec:discussion} and \ref{sec:conclusions}, respectively.
    
	\begin{table*}
		\caption{VLA1623 observational parameters}
		\label{tab:vla1623-obsparam}
		\begin{tabular}{ccccc}
			\hline
			Parameter name & Symbol & Value & Units & Reference\\
			\hline
			Protostellar mass of A1 & M$_{\mathrm{*,A1}}$ & 0.135$\pm$0.03 & M$_{\odot}$ & (1)\\
			Protostellar mass of A2 & M$_{\mathrm{*,A2}}$ & 0.135$\pm$0.03 & M$_{\odot}$ & (1)\\
			Protostellar mass of B & M$_{\mathrm{*,B}}$ & 1.9$\pm$0.25 &  M$_{\odot}$ & (1)\\
			Protostellar mass of W & M$_{\mathrm{*,W}}$ & 0.64$\pm$0.06 &  M$_{\odot}$ & (1)\\
			Circumbinary disk mass of A1+A2 & M$_{\mathrm{d,A}}$ & 54$\pm$5 & 10$^{-3}$ M$_{\odot}$ & (1)\\
			Disk mass of B & M$_{\mathrm{d,B}}$ & 1 & 10$^{-3}$ M$_{\odot}$ & (1)\\
			Disk mass of W & M$_{\mathrm{d,W}}$ & 1.7$\pm$0.2 & 10$^{-3}$ M$_{\odot}$ & (1)\\
			Inclination of A2 wrt A1 & $i_{\mathrm{A2}}$ & 58.9$\pm$0.45 & degrees & (1)\\
			Inclination of B wrt A1 & $i_{\mathrm{B}}$ & 91$\pm$10 & degrees & (1)\\
			Inclination of W wrt A1 & $i_{\mathrm{W}}$ & 74.8$\pm$0.5 & degrees & (1)\\
			Separation A1-A2 & $S_{\mathrm{A1-A2}}$ & 0.1 & $\arcsec$ & (1,2)\\
			Separation A1-B & $S_{\mathrm{A1-B}}$ & 1 -- 4 & $\arcsec$ & This work \\
			Separation A1-W & $S_{\mathrm{A1-W}}$ & 10.7 & $\arcsec$ & (1)\\
			\hline
			\multicolumn{5}{c}{Accretion and mass loss}\\
			\hline
			Accretion rate of A1 & $\dot{M}_{\mathrm{A1}}$ & 1.43$\pm$0.45 & 10$^{-8}$ M$_{\odot}$~yr$^{-1}$ & This work (Sect.~\ref{subsec:vla1623-in-out})\\
			Accretion rate of A2 & $\dot{M}_{\mathrm{A2}}$ & 1.43$\pm$0.45 & 10$^{-8}$ M$_{\odot}$~yr$^{-1}$ & This work \\
			Accretion rate of B & $\dot{M}_{\mathrm{B}}$ & 2.26$\pm$0.54 & 10$^{-9}$ M$_{\odot}$~yr$^{-1}$ & This work \\
			Accretion rate of W & $\dot{M}_{\mathrm{W}}$ & 3.49$\pm$0.42 & 10$^{-9}$ M$_{\odot}$~yr$^{-1}$ & This work \\
			Outflow mass loss rate & $\dot{M}_{\mathrm{loss}}$ & 2.3 & 10$^{-6}$ M$_{\odot}$~yr$^{-1}$ & (3)\\
			\hline
			\multicolumn{5}{c}{Cloud Core parameters}\\
			\hline
			Inner radius & $R_{\mathrm{in}}$ & 4.3 & au & (4)\\
			Outer radius & $R_{\mathrm{out}}$ & 10$^{4}$ & au & (4)\\
			Reference radius & $R_{\mathrm{ref}}$ & 10$^{3}$ & au & (4)\\
			Reference number density & $n_{\mathrm{ref}}$ & 7.7$\pm$1.54 & 10$^{5}$ cm$^{-3}$ & (4) \\
			Density profile index & $\alpha$ & -1.4 & dimensionless & (4)\\
			\hline
            \multicolumn{5}{l}{References: (1) \citet{sadavoy2024vla1623masses}; (2) \citet{harris2018A}; (3) \citet{andre1990}; (4) \citet{jorgensen2002}.}\\
		\end{tabular}
        
	\end{table*}

    \begin{figure*}
        \centering
        \includegraphics[width=0.95\linewidth]{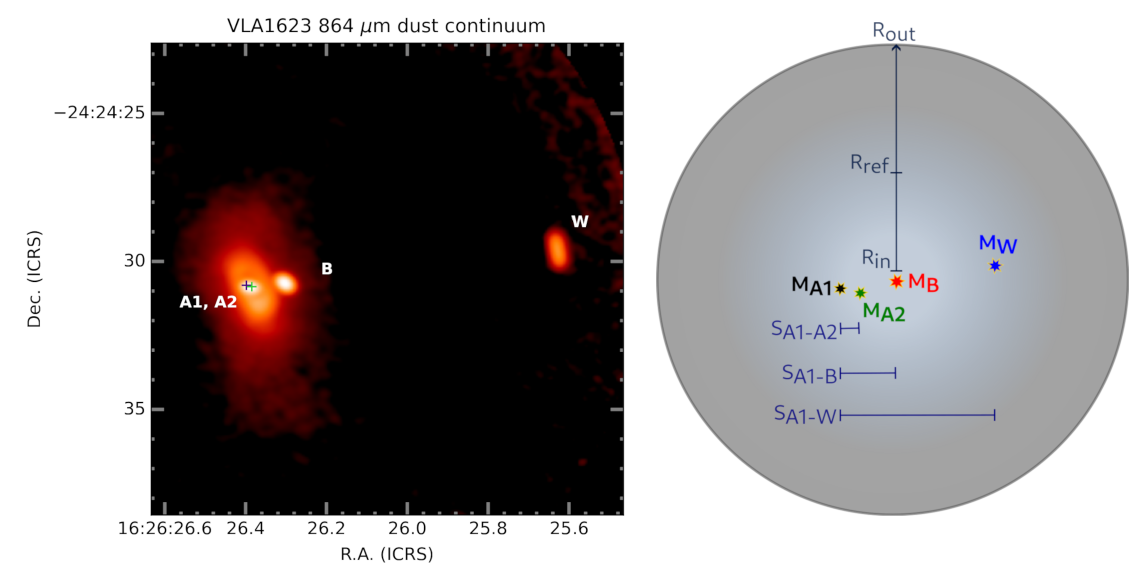}
        \caption{The quadruple protostellar system VLA1623. \textit{Left:} Dust continuum image of VLA1623 at 864 $\mu$m. The dust continuum peaks from left to right show the unresolved A1+A2 close binary, B and W. The large dust circumbinary disk is also visible in the continuum image. \textit{Right:} Cartoon showing the system setup in our simulations. The mass for each component $M_{p}$ in the simulation is the sum of the protostellar mass $M_{*,p}$ and gas disk masses $M_{d,p}$.}
        \label{fig:VLA1623_config}
    \end{figure*}

	\section{VLA1623 observational constraints}
    \label{sec:VLA1623data}
    The following subsections detail the parameters for the $N$-body simulations based on the observational constraints of VLA1623. An 864 $\mu$m dust continuum image of VLA1623 (ALMA project ID 2018.1.01089, image processing carried out via the NRAO ALMA pipeline \citealt{alma2023}) is shown in Figure~\ref{fig:VLA1623_config}. The adopted parameters are listed in Table~\ref{tab:vla1623-obsparam}.

	\subsection{Masses and inclinations}
	\label{subsec:vla1623-mass}
	
	For our simulations, we adopt the protostar $M_{*}$, gas disk masses $M_{d}$ and gas disk inclination $i$ as the orbital inclination, along with the corresponding uncertainties, for A, B, and W from \cite{sadavoy2024vla1623masses}.
	As a simplification, the mass for each protostar in the simulation $M_{p}$ is the sum of the corresponding disk mass $M_{d,p}$ and the protostellar mass $M_{*,p}$ for each source in the simulation.
	We assume $M_{A1}$/$M_{A2}$ = 1, hence distributing the protostellar and disk gas-mass of A evenly among the A1 and A2 components.

	\subsection{Component separations}
	\label{subsec:vla1623-sep}
    The adopted A1-A2 separation is 0.1$\arcsec$, considering the projected separation (14 AU) reported in \citet{sadavoy2024vla1623masses} which was based on \citet{harris2018A}.
    For the A1-W separation we adopt the projected separation of 10.7$\arcsec$ (e.g., \citealt{sadavoy2024vla1623masses,murillo-lai2013}).
    The uncertainty for both separations is taken from the adopted distance.
    
	The projected A1-B separation reported in literature, 1.1$\arcsec$ ($\sim$152 AU at $d\sim$138 pc), is smaller than the summed disk radii of both components without accounting for inclination (VLA1623A disk: 150-180 AU, \citealt{murillo2013}; VLA1623A disk: 316$\pm$32 AU, VLA1623B disk: 43$\pm$5 AU \citealt{sadavoy2024vla1623masses}).
    No substructure has been found in the disks either \citep{radley2025,maureira2025}.
    It is thus possible that the A1-B separation could be larger than 1.1$\arcsec$. Hence, the disk truncation radius limit presented in \citet{winter2018} is used to constrain the range of separations that can produce the observed disks of A (150--316 au) and B (43 au). The minimum separation can be obtained by:
	\begin{equation}
		R'_\mathrm{out} \approx 0.6~e_{\mathrm{pert}}^{0.11f}~\left(\frac{M_2}{M_1}\right)^{-0.2}~x_{\mathrm{min}}
	\end{equation}
	where $f = (\frac{M_2}{M_1})^{-1/3}$, $e_{\mathrm{pert}}$ is the eccentricity of the perturber, $M_{1}$ is the mass of the host, $M_{2}$ is the mass of the source passing by (perturber), and $x_{\mathrm{min}}$ is the minimum distance between components to produce the truncated radius $R'_{out}$.
	Considering either A or B as a perturber of the other, we find a range of 1$\arcsec$--4$\arcsec$, equivalent to 138--552 AU at $d\sim$138 pc for eccentricities smaller than 0.75.
	The lower end of the range corresponds to B being perturbed by A and having a truncated disk. The upper end of the range considers component B as the perturber while allowing the large circumbinary A disk.
	
	\subsection{Cloud core}
	\label{subsec:vla1623-cloudcore}
    VLA1623 is observationally found to be situated in a cold and dense cloud core \citep{jorgensen2002,bergman2011,chen2018,murillo2018SED,kawabe2018,radley2025}.
    This means that the cloud core has enough mass to present a relevant gravitational force on the components of VLA1623 throughout the protostellar phase, thus affecting the dynamical evolution of the system.
	The cloud core density profile is adopted from \citet{jorgensen2002}, which used single-dish observations of protostellar cloud cores together with 1D radiative transfer modeling. 
    The density profile is described by an inner and outer radii, $R_{\mathrm{in}}$ \& $R_{\mathrm{out}}$, power-law index $\alpha$, and reference density $n_{\mathrm{ref}}$ at a reference radius $R_{\mathrm{ref}}$.
    The cloud core mass obtained from this density profile is between 3 and 5 M$_{\odot}$, depending on the reference number density value used.
    The density profile has been previously used to model the molecular gas envelope of VLA1623 \citep{murillo2015,murillo2018}, and was found to reproduce the observed density and chemical conditions of the gas. Table \ref{tab:vla1623-obsparam} provides the cloud core adopted parameters.
	
	\subsection{Outflows and accretion}
	\label{subsec:vla1623-in-out}

    Outflows contribute to the mass loss and dispersal of the cloud core (e.g., \citealt{arce2006}).
    We include this effect as a single, constant mass loss rate adopted from \citet{andre1990}. By adopting an outflow mass loss rate calculated from observations that trace almost the full extent of the molecular gas outflow, we reduce possible bias caused by missing structure. 

    Mass accretion rates for each component in VLA1623 are calculated with the assumption used in \citet{codella2024}, where the bolometric luminosity is assumed to equal the accretion luminosity (i.e., $L_{\mathrm{bol}}$ = $L_{\mathrm{acc}}$). The accretion rate is then given by the expression:
    \begin{equation}
		\dot{M}_{\mathrm{acc,}p} = \frac{L_{\mathrm{bol,}p}R_{*}}{GM_{*,p}},
	\end{equation}
	where $L_{\mathrm{bol,}p}$ is the bolometric luminosity of protostar $p$, adopted from \citet{murillo2018SED}, $R_{*}$ is the protostellar radius set to 2$R_{\odot}$ \citep{codella2024}, $G$ is the universal gravitational constant, and $M_{*,p}$ is the protostellar mass of component $p$.
	We calculate mass accretion rates of 10$^{-8}$ M$_{\odot}$~yr$^{-1}$ for A1+A2, and an order of magnitude lower for B and W (Table~\ref{tab:vla1623-obsparam}). 
	Considering that A1+A2 are the most luminous components of VLA1623, this result makes sense.
	
    \subsection{Proper motion information}
	\label{subsec:vla1623-motion}
	There have been previous attempts to perform a proper motion analysis for the components in VLA1623.
	\citet{codella2024} used data from only two epochs (i.e., dust continuum images from \citet{harris2018A} and \citet{codella2024} with different angular resolutions) and found that A1+A2 and B have consistent motions. Their analysis does not provide constraints for W.
	In \citet{sadavoy2024vla1623masses}, a proper motion analysis was performed with data from six epochs (2013 to 2024), but was only done for components B and W, with no constraints for the proper motion of A (A1+A2).
    Due to the limitations of the available proper motion analysis we did not adopt these constraints into our simulations.

	\section{$N$-body simulations}
	\label{sec:simulations}
	\subsection{Set-up}
    \label{subsec:setup}
    We made use of the REBOUND $N$-body code \citep{rebound} to perform our simulations. 
	The simulations were integrated using IAS15, a 15th order Gauss-Radau integrator \citep{reboundias15}. 
    For each component $p$ in our REBOUND simulations we assign mass, semi-major axes, eccentricity, and inclination based on observational constraints (Table~\ref{tab:vla1623-obsparam}). 
    We adopt the formal value of the gravitational constant, which means the simulations are in units of kg, m, s, and m/s.
	Before integrating at each timestep, the system is moved to the center of mass (COM) reference frame.
	
	We performed two types of simulations: stellar-gravity-only, and with the addition of the gravitational force produced by the cloud core mass. In the latter type, the cloud core mass is gradually reduced by the combined effect of the outflow mass loss rate and the mass accretion rate onto the protostars.
	For the stellar-gravity-only simulations, only REBOUND is used. 
	In order to implement additional astrophysical effects for the simulations with the cloud core, we wrote two routines in REBOUNDx \citep{tamayo2020reboundx} to implement protostellar mass accretion onto each component, and adjust the component accelerations due to the presence of the cloud core. The implementation of these effects is described in the following paragraphs.
    
    The cloud core mass $M_{\mathrm{core}}$ at any timestep is calculated based on the adopted density profile with index $\alpha$ with the expression:
    \begin{equation}
    \label{eq:Mcore}
        M_{\mathrm{core}} = 4\pi A_{\mathrm{core}} \eta_{\mathrm{core}} (R_{out}^{\eta_{\mathrm{core}}} - R_{\mathrm{in}}^{\eta_{\mathrm{core}}}),
    \end{equation}
    where $R_{in}$ and $R_{out}$ are the inner and outer radii of the cloud core, respectively, $\eta_{\mathrm{core}}$ = 3 + $\alpha$, and $A_{\mathrm{core}}$ is given by
    \begin{equation}
    \label{eq:Acore}
        A_{\mathrm{core}} = \frac{n_{\mathrm{ref}}}{R_{\mathrm{ref}}^{\alpha}},
    \end{equation}
    where $n_{\mathrm{ref}}$ is the reference number density at radius $R_{\mathrm{ref}}$.
    At each timestep, the initial cloud core mass $M_{\mathrm{core,0}}$ is reduced by the sum of the $\dot{M}_{\mathrm{acc,}p}$ for each $p$ and the mass loss generated by the outflow $\dot{M}_{\mathrm{loss}}$. The cloud core $R_{in}$ and $R_{out}$ are kept constant, hence the mass reductions to the cloud core are given by
    \begin{equation}
        M_{\mathrm{core}} = M_{\mathrm{core,0}} - [(\dot{M}_{\mathrm{A1}} + \dot{M}_{\mathrm{A2}} + \dot{M}_{\mathrm{B}} + \dot{M}_{\mathrm{W}} + \dot{M}_{\mathrm{loss}}) \times dt ],
    \end{equation}
    where $dt$ is the timestep. The reference density $n_{\mathrm{ref}}$ is recalculated at each timestep using Equations~\ref{eq:Mcore} and \ref{eq:Acore}.
    The protostellar mass accretion effect is implemented by increasing the mass of each component $p$ at each timestep by $\dot{M}_{\mathrm{p}}$ $\times$ $dt$. This produces a constant mass accretion onto the components in the simulation over the lifetime of the cloud core.
    The cloud core mass, protostellar mass accretion and outflow mass loss rates are then set to zero when $M_{\mathrm{core}}$ $\leq$ 0.1 M$_{\odot}$. This value is used as an indicator of the end of the protostellar cloud core lifetime \citep{offner2025}.

    The gravitational force produced by the cloud core is applied to each component $p$ in the simulation through the modification of its corresponding acceleration vector.
    For our case study, we center the cloud core on VLA1623 B, which means that the acceleration vector of B is not modified.
    At each timestep the distance $r_{p}$ between the position of a component in the system $r_{p,i}$ and the center of the cloud core $r_{\mathrm{core},i}$, where $i$ = (x, y, z), is given by 
    \begin{equation}
        r_p = \sqrt{\sum (r_{p,i} - r_{\mathrm{core},i})^2} = \sqrt{\sum (dr_{p,i})^2}.
    \end{equation}
    The cloud core mass contained within $r_p$, $M_{\mathrm{core},r_p}$, is calculated using the expression
    \begin{equation}
        M_{\mathrm{core},r_p} = 4\pi A_{\mathrm{core}} \eta_{\mathrm{core}} (r_p^{\eta_{\mathrm{core}}} - R_{in}^{\eta_{\mathrm{core}}}).
    \end{equation}
    The gravitational acceleration $g_{r_{p,i}}$ acting on the component is obtained by
    \begin{equation}
        g_{r_{p,i}} = \frac{- G M_{\mathrm{core,r}}~dr_{p,i}}{r^3}.
    \end{equation}
    The acceleration of a component $a_{p}$ at each timestep is then modified by adding $g_{r_{p,i}}$ to the acceleration $a_{p,i}$. This routine is applied to the components in our simulation (A1, A2, and W) at each timestep as long as $M_{\mathrm{core}}$ $\geq$ 0.1 M$_{\odot}$.

	We generate three simulation sets: i) Gravity Only; ii) Cloud Core $t_{\mathrm{life}}$; and iii) Cloud Core $t_{\mathrm{life}}$~$\times$~0.5 ($t_{\mathrm{life}}$ reduced by half). In the latter two simulation sets, $t_\mathrm{life}$ is the cloud core lifetime.
    With the adopted observational constraints (Table~\ref{tab:vla1623-obsparam}), $t_{\mathrm{life}}$ is on the order of 1 -- 2 Myr.
	The sudden removal of the cloud core once it reaches 0.1 M$_{\odot}$ is not expected to produce abrupt changes in the system dynamics (e.g., \citealt{calovic2025}). 
	However, $t_{\mathrm{life}}$ is expected to affect the dynamics of the system, as the gravitational force due to the cloud core mass would be applied for a different interval of time. Hence, simulation set iii) explores how $t_{\mathrm{life}}$ reduced by half affects the survivability of a higher-order protostellar system. The shorter lifetime is achieved by increasing the outflow mass loss rate $\dot{M}_{\mathrm{loss}}$ by a factor of 2.

    Our simulations focus purely on the gravitational dynamics of the model. Consequently, physical processes that can impact the evolution of multiple protostellar systems such as gas dynamical friction (drag), accretion angle, outflow position angle, magnetic fields, rotationally-supported disks, and other gas and dust structures (e.g., \citealt{offner2023}) were not included due to the scope of this paper and the complexity of implementing them in N-body simulations.

	\subsection{Parameter space}
	\label{subsec:paramspace}

    We use a Monte Carlo (MC) algorithm with a uniform distribution (numpy \texttt{random.uniform}) to randomly generate 1000 input conditions for each simulation set based on the observational uncertainties of VLA1623 (Table~\ref{tab:vla1623-obsparam}). For the minimum and maximum values of the uniform distribution, we have adopted a $\pm$3$\sigma$ spread from the mean values listed in Table~\ref{tab:vla1623-obsparam}. 
    Only the cloud core parameters (inner $R_{\mathrm{in}}$ and outer $R_{\mathrm{out}}$ radii, reference radius $R_{\mathrm{ref}}$, and power-law index $\alpha$) and outflow mass loss rate $\dot{M}_{\mathrm{loss}}$ are kept constant for all input conditions.
	The input conditions are the same for the three simulation sets with the objective to directly compare the results.
	
	Eccentricity $e$ is one of the difficult parameters to constrain from current embedded protostellar system observations.
	We assume an eccentricity of $e$ = 0 (i.e., circular orbit) for A2 around A1 since the disk does not show indications of internal asymmetries \citep{radley2025,maureira2025}. 
	For the eccentricity of the B and W orbits, we adopt a few assumptions based on observations of higher-order stellar systems.
	Substantial eccentricities are observed in 2+2 quadruples, $e=0.33$ and 0.56 \citep{powell2025}, while \citet{han2022} find a maximum value of $e=0.709$ for the inner eclipsing binary in a $2+1+1$ hierarchy with an orbital period of 3 days. Similarly, \citet{volkov2021} reported an $e=0.71$ for their quadruple system with a $2+2$ hierarchy and an orbital period of 6.6 yr about the barycenter.
	The most eccentric brown dwarf in a binary system exhibits $e\sim0.78$ in its orbit around a one solar mass star \citep{henderson2024}.
	Disk truncation is particularly sensitive to eccentricity \citep{cox2017,winter2018,manara2019}. Hence we can use this as an additional constraint, especially for the A1-B eccentricity (Sect.~\ref{subsec:vla1623-sep}).
	\citet{manara2019} find eccentricities of up to $\sim 0.7$ when considering gas disk radii (about 2 times the dust disk radii) in their truncation analysis.
	Considering all the above observational constraints, we adopt a maximum eccentricity of $e=0.75$ in our simulations.
	
	\begin{table*}
		\caption{Multiplicity Diagnostic criteria}
		\begin{tabular}{ccc}
			\hline
			Degree of Multiplicity & Energy of B wrt COM & Total Energy Pair Criteria \\
			\hline
			Quadruple & $E_\mathrm{min,B}$ < 0.0 & ($E_\mathrm{min,A1-B}$, $E_\mathrm{min,A2-B}$, $E_\mathrm{min,W-B}$) < 0.0 \\
              & & ($E_\mathrm{min,A1-B}$, $E_\mathrm{min,W-B}$, $E_\mathrm{min,A1-A2}$) < 0.0 \\
             & & ($E_\mathrm{min,A2-B}$, $E_\mathrm{min,W-B}$, $E_\mathrm{min,A1-A2}$) < 0.0 \\
             & & ($E_\mathrm{min,A1-B}$, $E_\mathrm{min,W-B}$, $E_\mathrm{min,A2-W}$) < 0.0 \\
             & & ($E_\mathrm{min,A2-B}$, $E_\mathrm{min,W-B}$, $E_\mathrm{min,A1-W}$) < 0.0 \\
            \hline 
            Triple (one ejection) & $E_\mathrm{min,B}$ < 0.0 & ($E_\mathrm{min,A1-B}$, $E_\mathrm{min,W-B}$) < 0.0\\
             & & ($E_\mathrm{min,A2-B}$, $E_\mathrm{min,W-B}$) < 0.0\\
             & & ($E_\mathrm{min,W-B}$, $E_\mathrm{min,W-A1}$) < 0.0\\
             & & ($E_\mathrm{min,W-B}$, $E_\mathrm{min,W-A2}$) < 0.0\\
             & & ($E_\mathrm{min,A1-B}$, $E_\mathrm{min,A2-B}$) < 0.0\\
            \hline
            Binary pair & $E_\mathrm{min,B}$ < 0.0 or $E_\mathrm{min,B}$ > 0.0 & ($E_\mathrm{min,A1-B}$, $E_\mathrm{min,W-A2}$) < 0.0 \\
             & & ($E_\mathrm{min,A2-B}$, $E_\mathrm{min,W-A1}$) < 0.0 \\
             & & ($E_\mathrm{min,A1-A2}$, $E_\mathrm{min,W-B}$) < 0.0 \\
            \hline
            One Binary (two ejections) & $E_\mathrm{min,B}$ < 0.0 or $E_\mathrm{min,B}$ > 0.0 & $E_\mathrm{min,A1-B}$ < 0.0 \\
             & & $E_\mathrm{min,A2-B}$ < 0.0 \\
             & & $E_\mathrm{min,W-B}$ < 0.0 \\
             & & $E_\mathrm{min,A1-A2}$ < 0.0 \\
             & & $E_\mathrm{min,A1-W}$ < 0.0 \\
             & & $E_\mathrm{min,A2-W}$ < 0.0 \\
			\hline
		\end{tabular}
		\label{tab:EnergyCriteria}
	\end{table*}

\subsection{Integration time}
\label{subsec:inte_time}

Our objective is to determine the survivability rate of VLA1623 as a gravitationally-bound quadruple system into the main sequence.
    Our initial time ($t~=$ 0) is the current state of VLA1623 based on observations (Sect.~\ref{sec:VLA1623data}).
    To define for how long to integrate the simulations, we use a combination of estimated lifetimes from the literature.
	In \citet{enoch2009}, the total lifetime from early Class 0 to late Class I is about 0.54 Myr based on total and embedded young stellar object statistics of the populations from a number of solar neighborhood clouds observed with \textit{Spitzer}.
	The lifetime estimations from \citet{offner2025} note that in their simulations the protostellar core lifetime is 0.8 to 1.1 Myr. In this lifetime, the protostellar phase lasts 0.1 Myr after which the mass accretion falls below detection limit.
	In another approach using a half-life method to account for variations in evolutionary stage lifetimes, \citet{kristensen2018} notes that the Class II stage has a half-life of 2 Myr, which might not be true for all cases. 
	Looking at pre-main sequence sources, we obtain additional constraints.
	\citet{reipurth2024} observe a low-mass pre-main sequence quadruple located in a 3-6 Myr association.
	This is in agreement with studies looking at young clusters \citep{gezer2025,sanchez2024,smith2023,zari2019,briceno2019}.
	Combining all these constraints from observations and models, and considering our objective is to explore which systems reach the main sequence as gravitationally-bound systems, we set our integration time to 8 Myr, in 2 $\times$ 10$^{5}$ timesteps linearly spaced, distributed approximately as 1 Myr for the embedded protostellar phase (Class 0 and I), 2--3 Myr for the pre-main sequence phase (Class II+III), and an additional 4--5 Myr for the main sequence phase.

	\subsection{Simulation Diagnostics}
	To classify the results of the simulations, we use the total energy ($E$) as the main diagnostic tool. 
    The total energy of each pair of components is used to determine gravitational boundness and classify the resulting system according to its degree of multiplicity.
    The total energy $E$ of a component $p$ and reference component $p'$ (i.e., COM, primary, or another component) is given by 
    \begin{equation}
        E_{p,p'} = (\frac{1}{2} M_{p} v_{p}^2) - \frac{(G M_{\mathrm{sys}} M_{p})}{r_{p-p'}},
    \end{equation}
    where $v_{p}$ is the velocity of the component $p$, $M_{\mathrm{sys}}$ is the sum of the masses of all four protostellar components, and $r_{p-p'}$ is the distance between a given component $p$ and the reference component $p'$. In the simulation sets where the cloud core is present, $M_{\mathrm{sys}}$ includes $M_\mathrm{core}$, as long as it is $>$0.1 M$_{\odot}$, in addition to the protostellar masses.
    Six $E$ pair parameters (A1, A2, and W wrt B; A1 and A2 wrt W; and A2 wrt A1) are calculated at each timestep. 
    Additionally, the total energy of B wrt the COM is calculated.

    The first criterion used to classify the outcome of systems is the minimum total energy of B wrt the COM, $E_\mathrm{min,B}$, in 8 $\times$ 10$^{4}$ yr (0.08 Myr, 2000 timesteps) intervals. 
    The reason for using this criteria is that the system COM is at the center of the reference frame (see Sect.~\ref{subsec:setup}).
    Thus, a bound system will have negative $E_\mathrm{min,B}$, while an ejection or perturbation in the system (e.g., reorganization of the system hierarchy) can cause $E_\mathrm{min,B}$ to become positive due to B moving away from the COM.

    To determine the temporal evolution of the degree of multiplicity, the minimum value $E_{\mathrm{min}}$ for each pair in 0.08 Myr intervals is obtained.
    The $E_\mathrm{min}$ value is used since the energy of any pair in higher-order multiples oscillates between negative and positive values even when bound. This effect is produced by the gravitational perturbations caused by the other components in a system.
    When $E_{\mathrm{min}}$ < 0, the pair is bound. Instead, when $E_{\mathrm{min}}$ $\geq$ 0, one of the components in the pair has been ejected.
    Table~\ref{tab:EnergyCriteria} lists the criteria used to determine the degree of multiplicity of the simulated system at each timestep.

    \begin{figure*}
	    \centering
        \includegraphics[width=0.95\linewidth]{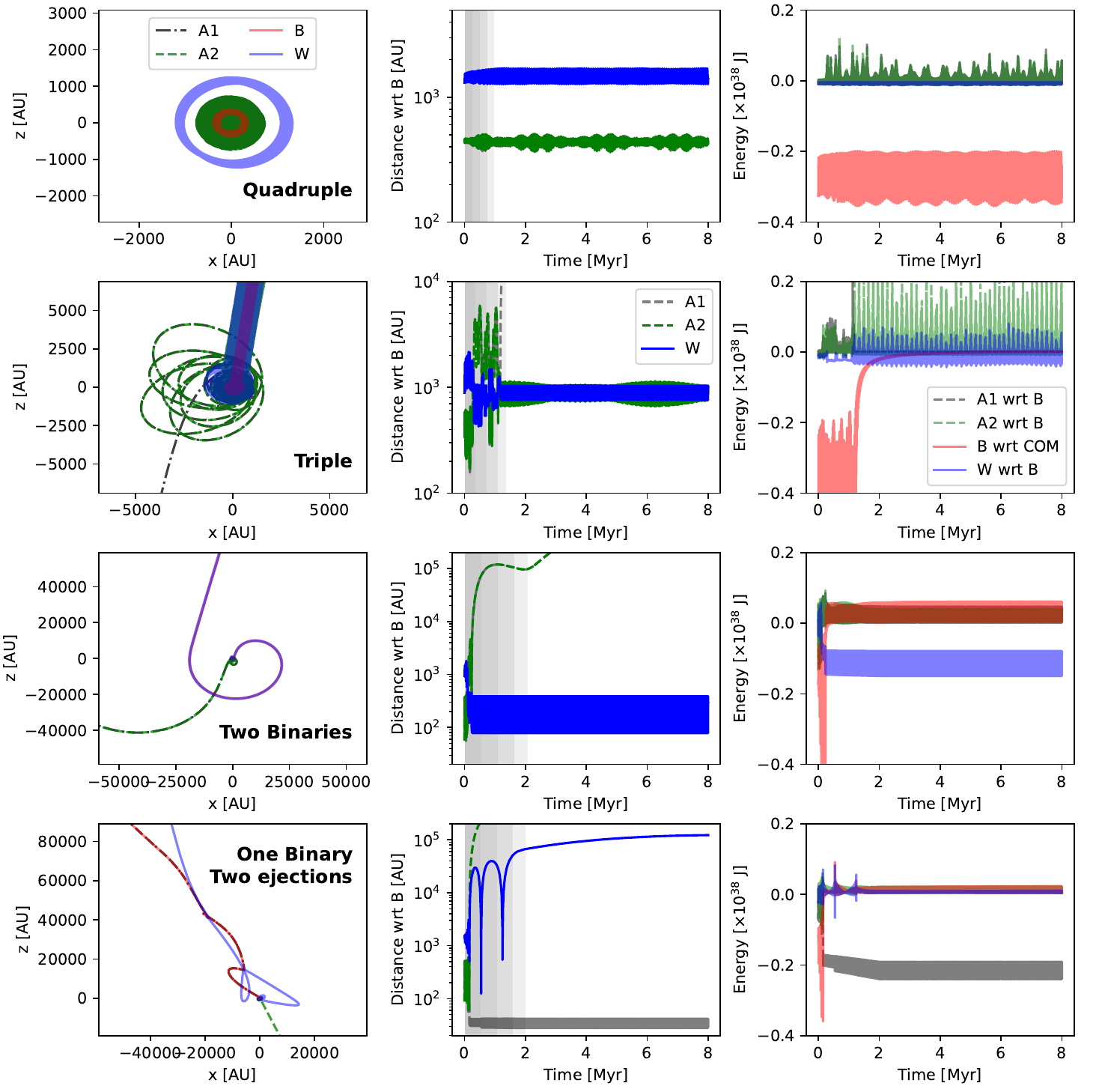}
        \caption{Representative simulations of the four possible outcomes for VLA1623 based on the simulations presented in this work. Results from the simulation set Cloud Core $t_\mathrm{life}$ are shown. The first column shows the orbits in the x-z plane. The second column shows the evolution of the distance of A1, A2 and W wrt B over the 8 Myr of the simulation. The gray shaded areas show, from darker to lighter, the duration of the cloud core at $\geq$0.75 mass, $\geq$0.5 mass, $\geq$0.25 mass and up to the point where $M_{core}$ $\leq$0.1 M$_{\odot}$. The third column shows the pair energy evolution over the 8 Myr of the simulation. In all plots, components A1, A2, B and W are shown with black, green, red and blue lines, respectively.}
        \label{fig:Orbits}
	\end{figure*}

    \begin{figure*}
	    \centering
        \includegraphics[width=0.85\linewidth]{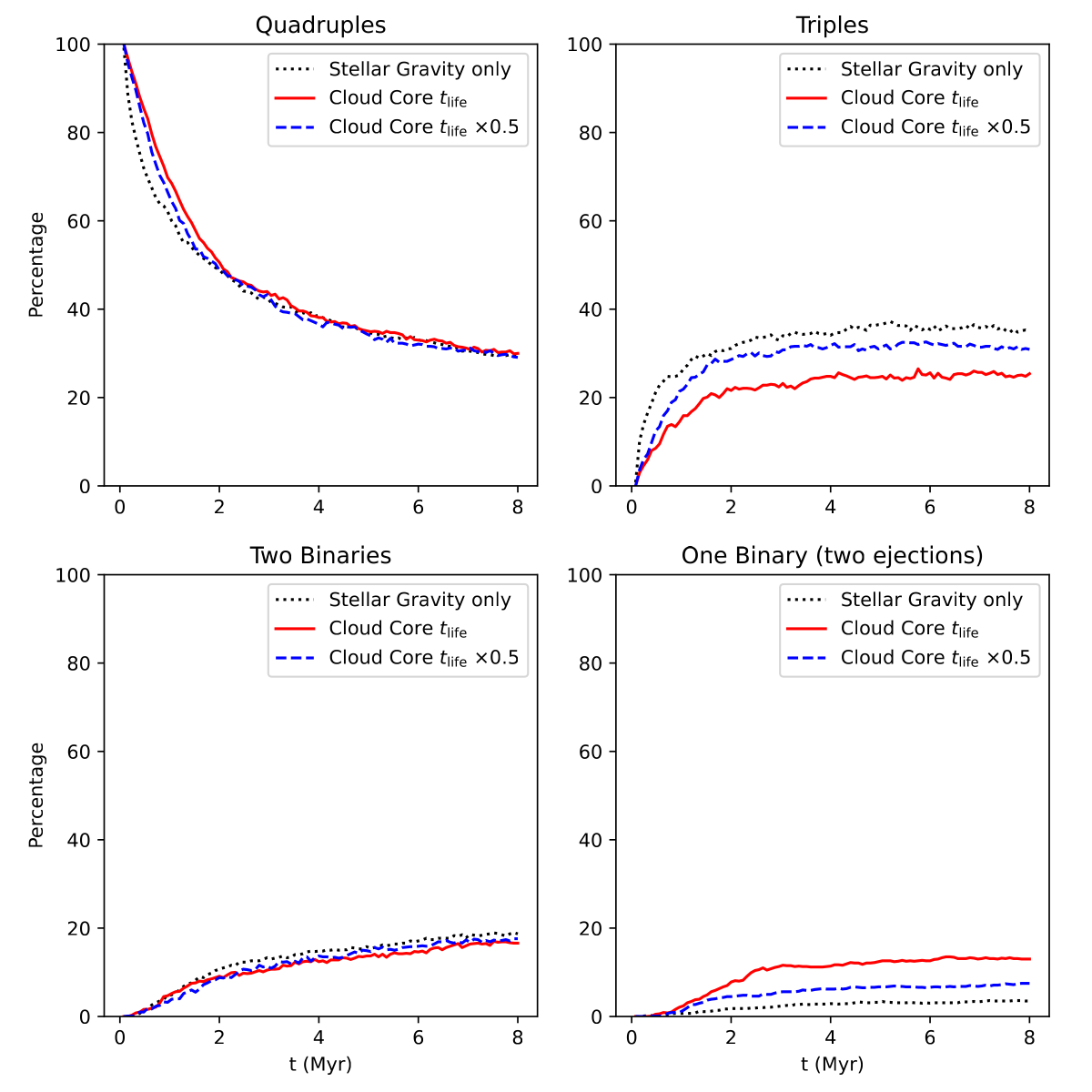}
        \caption{Percentage over time of quadruples (\textit{top left}), triples (\textit{top right}), two binaries (\textit{bottom left}), and one binary with two ejections (\textit{bottom right}). The red solid, blue dashed, and black dotted lines show the different simulation sets.}
        \label{fig:dtHistogram}
	\end{figure*} 
    
	\begin{table*}
		\caption{Classification of bound systems from the final 0.08 Myr of the simulations}
		\begin{tabular}{ccccccc}
			\hline
			Simulation & Quadruples & Triples & Pair Binaries & One Binary & Unclassified & Total non-quadruple  \\
			& \% & \% & \% & \% & \% & \% \\
			\hline
			Gravity only	&	30.0	&	34.9	&	18.7	&	3.5 &	12.9	&	70.0	\\
			Cloud Core $t_{\mathrm{life}}$ &	30.0	&	25.4	&	16.6	&	13.0	&	15.0	&	70.0	\\
			Cloud Core $t_{\mathrm{life}}$ $\times$ 0.5	&	29.1	&	30.9	&	17.6	&	7.5	&	14.9	&	70.9	\\
			\hline
		\end{tabular}
		\label{tab:classification}
	\end{table*}

    \section{Gravitationally-bound cases}
	\label{sec:results}
	For all three simulation sets, four possible outcomes occur: bound quadruple, bound triple with one ejected component, two binaries, and one binary with two ejected components. Four examples from the simulation set Cloud Core $t_\mathrm{life}$ are shown in Figure~\ref{fig:Orbits}.
    Unclassified systems are those whose dynamical behavior within a given time bin does not allow its classification with the energy pair criteria used in this work. Unclassified systems include the cases where all energy pairs are positive. This can indicate a dissolved system, but can also occur due to energy oscillations leading to all energy pairs being positive within a given time bin while the system is still a quadruple, triple or binary.
        
    Figure~\ref{fig:dtHistogram} shows how the percentage of quadruples, triples and binaries changes, in 0.08 Myr bins, over the total integration time of 8 Myr for the parameter space of VLA1623. 
    All configurations of VLA1623 start as quadruples in the simulations. During the first 2 Myr, the presence of the cloud core, and its $t_\mathrm{life}$, has a notable effect on the degree of multiplicity of the system. When compared to the gravity only simulations, the cloud core helps keep a higher percentage of quadruple systems bound during the protostellar phase and into the early pre-main sequence phase. After 2 Myr, the percentage of bound quadruples drops down to $\sim$45\%. Over the remaining 6 Myr, the three simulation sets show practically the same trend, and the percentage of quadruples drops to $\sim$30\% by the end of the three simulation sets.
    The percentage of triples rises to above 20\% within the first 0.5 Myr in the stellar-gravity-only simulation set. When a cloud core is present, the percentage of triples rises much slower and varies with $t_\mathrm{life}$.
    The formation of binaries, either two pairs or one pair with two ejections, gradually increases with time. However, a cloud core with a longer $t_\mathrm{life}$ notably increases the probability of one binary and two ejections.
    The resulting statistics per simulation set based on the final 0.08 Myr of the simulations are listed in Table~\ref{tab:classification}.

    The gravitationally-bound quadruples mainly have hierarchies of 1+2+1, meaning that B is ``closely'' orbited by the A1+A2 pair at a few 10$^{2}$ AU, while W orbits further out on the order of 10$^{3}$ AU. This configuration creates considerable oscillations in the total energy of A1 and A2, and can lead to unstable orbits.
    Such oscillations are consistent with the results from previous simulations \citep{offner2023}.
    It is not uncommon in the simulations for A1 and/or A2 to migrate to a much wider orbit on the order of 10$^{4}$ AU while still remaining a quadruple system with hierarchies of 2+2.

    Bonafide triple and binary systems are produced when one or two components are ejected from the system.
    Once a component is ejected, the remaining sources tighten the orbit, as expected.
    For triples, typically A1 or A2 are ejected at some point between 0.1 and 3 Myr. Ejections can occur early in the simulation before the cloud core mass is below 75\% of its initial mass (Fig.~\ref{fig:Orbits}, bottom row), or can occur as the cloud core disperses and has less than 25\% of its original mass (Fig.~\ref{fig:Orbits}, right column).
    Component W can be ejected when a triple forms, however it is more often ejected when two ejections occur and a binary remains.
    The primary B can also be ejected when one binary forms, but this occurs even less frequently.

    We constrain the simulation results with the current best fit protostellar masses (Table~\ref{tab:vla1623-obsparam}) to explore the range of possible A-B separations that lead to the survival of VLA1623 as a gravitationally-bound quadruple system.
    For A-B separations of 1.1" (current projected separation), VLA1623 can survive as a gravitationally bound quadruple system with $e<$0.4 for the A-B and B-W orbits.
    Increasing the A-B separation requires the eccentricity of A-B and B-W to tend toward 0 so that VLA1623 can remain as a gravitationally-bound quadruple system.

	\section{Discussion}
	\label{sec:discussion}

    In our simulations, we assume as an initial condition that the components A1, A2, B, and W are gravitationally bound. After 8 Myr, we find that VLA1623 has a 30\% chance to survive as a gravitationally bound quadruple system into the main sequence considering the current observational constraints. The possibility that VLA1623 dissolves into a triple system is 25 -- 35\%, which suggests that quadruple protostellar systems provide an additional pathway to form stable triple systems that survive into the main sequence.
    
    Our results show that the survivability and stability of VLA1623 depends on the combined configuration of protostellar mass ratios wrt the primary, separations between components, and eccentricities.
    Separations alone cannot determine if a higher-order multiple protostellar system will remain gravitationally-bound or if the protostellar system has undergone dynamical evolution.
    Component masses and separations can be derived from rotationally-supported gas disk analysis and dust continuum observations.
    On the other hand, eccentricities are harder to constrain. Possible ways could be through the modeling of gas and dust disk interactions between components (e.g., \citealt{winter2018,cuello2019}), or proper motion analysis.

    An interesting result from our simulations is that component W is the least likely to be ejected from the system. 
    This is surprising considering that W is located at a projected distance on the order of 10$^{3}$ AU and a different systemic velocity wrt the other components in the system \citep{murillo2013,mercimek2023}.
    If the inward migration of the wide separation components predicted from simulations including drag is also considered (\citealt{offner2023} and references therein), component W should not be ejected at all.
    Indeed, component W could in fact spiral inward due to the slow down caused by accretion \citep{kuruwita2023}.
    On the other hand, A1 and/or A2 are more often ejected in our simulations.
    Components A1 and A2 have the lowest masses in the VLA1623 system, thus it is not surprising that they have a larger probability of being ejected. 
    A burst of mass accretion onto these sources, effectively increasing their mass, could reduce the probability of their ejection and perturbations caused by the presence of component W.

    Ejected components from higher-order protostellar systems are usually considered to be one of the formation mechanisms of brown dwarfs (e.g., \citealt{reipurth2000,reipurth2010,reipurth2015,umbreit2005,umbreit2011}). 
	For the case study of VLA1623, all components are above the brown dwarf threshold of 0.08 M$_{\odot}$ (Table~\ref{tab:vla1623-obsparam}). 
	Thus, component ejections from VLA1623 will not lead to the formation of brown dwarfs, instead only very low-mass stars that could be captured by nearby protostars in the surrounding clustered environment of VLA1623 \citep{kuruwita2023,nakamura2025}.

    In the extreme cases where the quadruple system has a very stable, gravitationally-bound configuration, or an unstable and unbound configuration, the cloud core does not have a significant effect. The system will remain bound or become unstable regardless of the presence and lifetime of the cloud core.
    In contrast, the cloud core plays an important role in quadruple systems with configurations that are on the stability boundary. In these cases, the cloud core can provide an additional force to help make the system gravitationally-bound. The cloud core lifetime will matter in this process as well (Fig.~\ref{fig:dtHistogram}, top left panel).
    Around 1 Myr, there are 8\% (4\%) more gravitationally-bound quadruple protostellar systems in the Cloud Core $t_\mathrm{life}$ ($t_{\mathrm{life}}$~$\times$~0.5) simulations than in the stellar-gravity-only simulations.
    Additionally, the cloud core contributes to the formation of long-period gravitationally-bound quadruple systems with at least one component at separations on the order of 10$^{4}$ AU.
    The cloud core also serves to keep ejected components around for a fraction of the cloud core lifetime (Fig.~\ref{fig:Orbits}, bottom rows). This could impact the capture of weakly bound protostars \citep{kuruwita2023} in the clustered environment where VLA1623 forms \citep{nakamura2025}.
    Even more interesting is the effect the cloud core lifetime has on systems where individual components are ejected (Fig.~\ref{fig:dtHistogram}, left column). About 5\% less triple systems are formed when the cloud core lifetime is on the order of 1 to 2 Myr. In contrast, the probability of forming a binary with two ejected components increases by about a factor of 2 with longer cloud core lifetimes.

    The above noted effects can be better understood as an impulse (an additional central force being present over a period of time). Higher-order protostellar systems are typically formed and found in cloud cores with higher densities and masses than those of single or binary protostars \citep{luo2022,murillo2024,murillo2025}. Such cores tend to be located in regions of molecular clouds where velocity-coherent gas structures intersect or join (e.g., \citealt{hacar2017,chen2020,chen2024}), likely feeding these cloud cores and effectively extending their lifetime.
    The VLA1623 cloud core (Sect.~\ref{subsec:vla1623-cloudcore} \& \ref{subsec:vla1623-in-out}) has a current mass of 3--5 M$_{\odot}$, based on the density profile from \citet{jorgensen2002}, and it is located south from a cold and dense ridge of material \citep{bergman2011,chen2018}. A cloud core of 3.5 M$_{\odot}$ would take about 1 Myr to disperse down to 5\% of the current mass with accretion rates assuming $L_\mathrm{bol}$ = $L_\mathrm{acc}$ \citep{codella2024} and the derived outflow mass loss rate from observations \citep{andre1990}. Changes to the mass loss and accretion rates will change the VLA1623 cloud core lifetime.
    Hence, the cloud core lifetime will be a balancing act. Several factors will contribute to this process, including the amount of mass, the nature of accretion rates (steady vs. episodic), the uniformity of component accretion, and the number and direction of outflows capable of dispersing the cloud core (whether combined vs. individual, or aligned vs. orthogonal).

    We highlight that the present simulations do not consider drag due to the dense and cold gas \citep{bergman2011,chen2018} or breaking by the magnetic fields \citep{hull2014,harris2018A,sadavoy2018,kwon2018}, both relevant for VLA1623. 
	The directions of the outflows, direction of mass accretion from the cloud core or the disk, and the torques produced by the rotationally-supported disks are also not being factored into the simulations presented here.
	Thus, magneto-hydrodynamical (MHD) simulations are needed to determine the extent to which all the above factors and processes affect the survivability rate of VLA1623.
    Finally, our simulations are modeling an isolated higher-order protostellar system, when in fact VLA1623 is located in a clustered environment (e.g., \citealt{nakamura2025}).

    Previous studies suggest that the presence of drag from the cloud core, and gas accretion onto protostars would lead to an inward migration of components from 1000 AU to separations one or two orders of magnitude smaller within 1 Myr (\citealt{offner2023}, and references therein; \citealt{kuruwita2023}). 
    Our simulations include a cloud core and gas accretion onto all protostars, but no drag.
    The effect of drag could further tighten the stable quadruple protostellar systems by causing component W to migrate inward, or tighter orbits to form between A1+A2 and B.
    Another interesting point to explore would be how the A1+A2 circumbinary disk helps or hinders its membership in the VLA1623 system.
	In \citet{offner2023} and references therein, it is noted that the combination of equal mass protostars together with the circumbinary and circumstellar disks, ongoing accretion plus other disk parameters can actually lead to outward migration of the protostars in the close binary.
	VLA1623 A1 and A2 have identified individual dust disks \citep{radley2025,maureira2025}, and a rather massive circumbinary gas disk \citep{sadavoy2024vla1623masses,codella2022}. 
	Additionally, the pair is most likely accreting given its high luminosity relative to B and W \citep{murillo2018SED}. 
    Finally, it would be interesting to explore if the magnetic fields identified in and around the disk \citep{harris2018A,sadavoy2018,hull2014} play a role in keeping the binary bound.
	Based on the results from literature and our work, it may not be unlikely for A1 or A2 to migrate outward and either lead to a redistribution of the hierarchy or be ejected from the system.

	\section{Conclusions}
	\label{sec:conclusions}
	This work presents a series of $N$-body simulations aimed at exploring the survivability of higher-order protostellar systems ($\geq$3 components) with the case study of VLA1623, where we assumed that the quadruple  protostellar system is gravitationally bound.
	Current observational constraints of the masses, component separations, cloud core properties, protostellar mass accretion rates, and outflow mass loss rates were used to generate a thousand input conditions using a MC algorithm. The input conditions were then run through three sets of simulations: i) stellar-gravity-only; ii) Cloud Core $t_\mathrm{life}$; and iii) Cloud core $t_\mathrm{life}$ $\times$ 0.5 (cloud core lifetime reduced by half). Each simulation was integrated over 8 Myr, assuming roughly 1 Myr for the protostellar phase, 2-3 Myr for the pre-main sequence phase, and 4-5 Myr for the main sequence phase.
	The simulation outcomes were classified based on the energy of component pairs and of B wrt COM, with negative minimum total energies signaling gravitationally-bound pairs. Four outcome scenarios were found: bound quadruples, triples, a pair of binaries, and one binary with two ejected components. About 15\% of the systems are unclassified from their total energies due to the complexity of the configurations.
	
	Based on this work, we report the following conclusions:
	\begin{enumerate}
		\item The probability of VLA1623 surviving into the main sequence as a gravitationally-bound quadruple system is 30\%. This probability is consistent in all the simulation sets. Stable gravitationally-bound configurations depend on a combination of mass ratios wrt the primary, separations, and eccentricities. No single parameter dominates the rate of survivability of VLA1623.
		\item The probability of VLA1623 dissolving into a triple system (i.e., ejection of one component) is between 25 and 35\%. This outcome suggests that quadruple protostellar systems can contribute significantly to the production of stable triple (proto)stellar systems. Longer cloud core lifetimes reduce the probability of forming a triple system. Outcomes where a pair of binaries are formed (17--19\%) or a single binary remains with two ejections (4--13\%) are less common.
		\item Component VLA1623 W is the least likely to be ejected. In contrast, A1 or A2 are the most commonly ejected components. Addition of gas dynamical friction in the simulations may alter this outcome.
		
	\end{enumerate}
	
	The case study of VLA1623 illustrates the importance of a global view when studying the evolution of higher-order protostellar systems. Both the gas and the dust need to be observed and characterized in order to study the evolution of higher-order multiples. Protostellar masses, presence of disks, separation constraints, cloud core density profile, accretion and mass loss rates, and distribution of material among components will provide crucial information to determine the survivability and stability of higher-order protostellar systems.
	
	\section*{Acknowledgements}
	
	We thank the anonymous referee for its careful review that helped us to improve this work. N.M.M. thanks Shuo Huang, Chris Ormel, and Shoji Mori, for comments during a visit to Tsing Hua University which contributed to the initial phase of this work. We thank Stefano Souza for the discussion about the energy criteria method.
    This paper made use of the following ALMA data: ADS/JAO.ALMA 2018.1.01089. ALMA is a partnership of ESO (representing its member states), NSF (USA), and NINS (Japan), together with NRC (Canada) and NSC and ASIAA (Taiwan), in cooperation with the Republic of Chile. The Joint ALMA Observatory is operated by ESO, AUI/NRAO, and NAOJ.
	N.M.M. acknowledges support from the DGAPA–PAPIIT IA103025 grant. A.P.V. acknowledges support from the DGAPA–PAPIIT IA103224 and IN112526 grants.

	\section*{Data Availability}
	The model grids are available upon request to the authors.

	
	
	\bibliographystyle{mnras}
	\bibliography{MPSdynamics} 

	


	\bsp	
	\label{lastpage}
\end{document}